# Lid-driven cavity flow of viscoelastic liquids

R G Sousa [1], R J Poole[2], A M Afonso[1], F T Pinho[3], P J Oliveira[4], A Morozov[5] and M A Alves[1]✉

[1]Dep. of Chemical Eng., CEFT, Faculty of Engineering, University of Porto, 4200-465 Porto, Portugal

[2]School of Eng., University of Liverpool, Liverpool L69 3GH, UK

[3] Dep. of Mechanical Eng., CEFT, Faculty of Engineering, University of Porto, 4200-465 Porto, Portugal

[4] Dep. Eng. Electromecanica, C-MAST Unit, Universidade da Beira Interior, 6201-001 Covilhã, Portugal

[5] School of Physics and Astronomy, University of Edinburgh, Edinburgh EH9 3JZ, UK

✉Corresponding author: mmalves@fe.up.pt

## Abstract

The lid-driven cavity flow is a well-known benchmark problem for the validation of new numerical methods and techniques. In experimental and numerical studies with viscoelastic fluids in such lid-driven flows, purely-elastic instabilities have been shown to appear even at very low Reynolds numbers. A finite-volume viscoelastic code, using the log-conformation formulation, is used in this work to probe the effect of viscoelasticity on the appearance of such instabilities in two-dimensional lid-driven cavities for a wide range of aspect ratios (0.125 ≤ Λ=height/length ≤ 4.0), at different Deborah numbers under creeping-flow conditions and to understand the effects of regularization of the lid velocity. The effect of the viscoelasticity on the steady-state results and on the critical conditions for the onset of the elastic instabilities are described and compared to experimental results.

# 1      Introduction

The fluid motion in a box induced by the translation of one wall – the so-called "lid-driven cavity flow" – is a classic problem in fluid mechanics [1]. The geometry is shown schematically in **Figure 1** and conventionally comprises a two-dimensional rectangular box of height $H$ and width $L$ of which the top (horizontal) wall – the "lid" – translates horizontally at a velocity $U$ (for single phase fluids flowing isothermally the exact choice of moving wall is unimportant). For Newtonian fluids in rectangular boxes, the problem is governed by two dimensionless parameters: the Reynolds number $Re$ ($\equiv \rho UH/\eta$) and the aspect ratio $\Lambda$ ($\equiv H/L$), where $\rho$ is the fluid density and $\eta$ is the dynamic viscosity. For very low Reynolds numbers – creeping-flow conditions, where $Re \to 0$ – and low aspect ratios ($\Lambda$<1.6 [2]) a main recirculating region of fluid motion is induced which, due to the linearity of the Navier-Stokes equations under Stokes flow conditions, is symmetric about the vertical line $x/L$=0.5, as shown in **Figure 1**. In addition to this main recirculation, smaller Moffatt [3] or corner eddies are also induced at the bottom corners (labelled "C" and "D" in **Figure 1**): in fact at the bottom corners there is an infinite series of these vortices of diminishing size and strength as the corner is approached [3]. The primary corner eddies grow in size with increasing aspect ratio and, at a critical aspect ratio of about 1.629 [2], merge to form a secondary main cell albeit of much smaller intensity than the primary main cell (Ref. [1] gives a stream function decay ratio of 1/357 for these main cells). For higher aspect ratios this process repeats and additional main cells are created. Thus for "high" aspect ratios, essentially $\Lambda$>1.6, the main fluid motion near the translating wall (the main cell) is essentially unaffected by the aspect ratio. For a fixed aspect ratio – the majority of studies use $\Lambda$=1 – increasing the Reynolds number firstly breaks the fore-aft symmetry about the vertical line $x/L$=0.5 and then simulations reveal increasing complexity [1, 4]. Since experiments show [5] that above a critical Reynolds number of around 500 the flow becomes three-dimensional and then time-dependent ($Re \approx 825$), we will not discuss these high $Re$ steady two-dimensional simulations further here. Given the geometrical simplicity, combined with the rich observable fluid dynamics, the lid-driven cavity became a benchmark problem in the (Newtonian) fluid mechanics community for the development and validation of numerical schemes and discretization techniques [1,4,6,7].

For viscoelastic fluids the literature is understandably less dense but, at least under creeping-flow conditions, much has been revealed by the limited number of studies to date. Fluid viscoelasticity introduces two additional non-dimensional parameters to the problem: the Deborah number ($De$) which is defined as the ratio of the fluid's relaxation time ($\lambda$) to a characteristic residence time of the flow (which can be estimated as $L/U$) and the Weissenberg number ($Wi$) which is defined as the ratio of elastic ($\propto \lambda\eta U^2/H^2$) to viscous stresses ($\propto \eta U/H$). Therefore $De=\lambda U/L$ and $Wi=\lambda U/H$. Thus only in the unitary aspect ratio case are the two definitions identical: otherwise they are related through the aspect ratio ($De=\Lambda Wi$). Experimentally, the papers of Pakdel and co-workers [8-10] detailed the creeping flow of two constant-viscosity elastic liquids (known as Boger fluids [11]) – dilute solutions of a high molecular weight polyisobutylene polymer in a viscous polybutene oil – through a series of cavities of different aspect ratios ($0.25 \leq \Lambda \leq 4.0$). Pakdel et al. [9] initially characterized the flow field at low wall velocities where the base flow remained steady and approximately two-dimensional: viscoelasticity was seen to break the fore-aft symmetry observed in Newtonian creeping flow and the eye of the recirculation region moved progressively further to the upper left quadrant (i.e. towards corner A of **Figure 1**) of the cavities (incidentally, the fore-aft symmetry breaking due to inertia moves the eye towards corner B). At higher wall velocities, it was found [8,10,12] that the flow no longer remains steady but becomes time dependent. As inertia effects are vanishingly small in these highly-viscous Boger fluid flows, Pakdel and McKinley [10] associated this breakdown of the flow to a purely-elastic flow instability. The effect of cavity aspect ratio on this critical condition was found experimentally to occur at an approximately constant Deborah number (or, equivalently, $1/Wi$ scaling linearly with aspect ratio $\Lambda$). Pakdel and McKinley [8] were able to explain this dependence on aspect ratio via a coupling between elasticity and streamline curvature and proposed a dimensionless criterion (now often referred to as the "Pakdel-McKinley" criterion [13-15]) to capture this dependence on aspect ratio for both the lid driven cavity [8] and for a range of other flows [12].

Grillet et. al. [16] used a finite element technique to compute the effect of fluid elasticity on the flow kinematics and stress distribution in lid driven cavity flow, with a view to better understand the appearance of purely elastic instabilities in recirculating flows. In an effort to mimic the experiments and reduce, or circumvent, the numerical problems associated with the presence of a corner between a moving wall and a static one, the corner singularities were treated by incorporating a controlled amount of leakage. The results captured the experimentally observed upstream shift of the primary recirculation vortex and, concerning elastic instabilities, a dual instability mechanism was proposed, depending on aspect ratio (see also [17]). For shallow aspect ratios, the downstream stress boundary layer is advected to the region of curvature at the bottom of the cavity resulting in a constant critical Weissenberg number; in deep cavities, the upstream stress boundary layer is advected to the region of curvature near the downstream corner (B in **Figure 1**), resulting in a constant critical Deborah number.

Much as has been done for Newtonian fluids [6], a number of studies have used the lid-driven cavity set-up to numerically test novel approaches or benchmark codes for viscoelastic fluids. These include Fattal and Kupferman [18], who first proposed the log-conformation approach, and Pan et al. [19], Habla et al. [20], Comminal et al. [21], and Martins et al. [22], who later used that approach under the original or a modified form. Recently Dalal et al. [23] analysed the flow of shear thinning viscoelastic fluids in rectangular lid-driven cavities, but also used the Oldroyd-B model for validation of the code. All these authors have applied a 4th order polynomial velocity regularization (what we will refer to later in Section 3 as "R1") to simulate the Oldroyd-B flow in 2D lid-driven cavities (Ref. 22 has tackled the 3D flow). In **Table 1** we provide an overview of these previous numerical studies including the constant-viscosity viscoelastic model used (primarily Oldroyd-B with solvent-to-total viscosity ratio $\beta$=0.5), the numerical and regularization methods used and the Weissenberg number reached. An exception was Yapici et al. [24] who solved the Oldroyd-B model for $0 \leq Wi \leq 1$ with a finite-volume method, using the first-order upwind approximation for the viscoelastic stress fluxes in the rheological equation, and without recourse to the log-conformation approach. In marked contrast to all other studies with constant-viscosity viscoelastic models, including the present one, Yapici et al. [24] claim to be able to simulate viscoelastic lid-driven cavity flow without recourse to wall regularization.

Not surprisingly, there are a number of other studies for other types of non-Newtonian fluids, e.g. concerned with viscoplastic fluids where the interest is to identify the un-yielded central region (Mitsoulis and Zisis [25], Zhang [26], amongst others), and we shall use their results, in the Newtonian limit, as a basis for comparison.

In this work we re-visit the lid-driven cavity flow of viscoelastic fluids and investigate in detail how the choice of velocity regularization affects the viscoelastic simulations and the critical conditions under which the purely-elastic instability occurs.

## 2     Governing equations and numerical method

We are concerned with the isothermal, incompressible flow of a viscoelastic fluid flow, and hence the equations we need to solve are those of conservation of mass

$$\nabla \cdot \mathbf{u} = 0 , \tag{1}$$

and of momentum

$$\rho \frac{\partial \mathbf{u}}{\partial t} + \rho \mathbf{u} \cdot \nabla \mathbf{u} = -\nabla p + \nabla \cdot \boldsymbol{\tau}, \tag{2}$$

together with an appropriate constitutive equation for the extra-stress tensor, $\boldsymbol{\tau}$. In the current study we use both the upper-convected Maxwell (UCM) and Oldroyd-B models [27]

$$\boldsymbol{\tau} = \boldsymbol{\tau}_s + \boldsymbol{\tau}_p, \tag{3.1}$$

$$\boldsymbol{\tau}_s = \eta_s \left( \nabla \mathbf{u} + \nabla \mathbf{u}^T \right), \tag{3.2}$$

$$\boldsymbol{\tau}_p + \lambda \left( \frac{\partial \boldsymbol{\tau}_p}{\partial t} + \mathbf{u} \cdot \nabla \boldsymbol{\tau}_p \right) = \eta_p \left( \nabla \mathbf{u} + \nabla \mathbf{u}^T \right) + \lambda \left( \boldsymbol{\tau}_p \cdot \nabla \mathbf{u} + \nabla \mathbf{u}^T \cdot \boldsymbol{\tau}_p \right), \tag{3.3}$$

where $\lambda$ is the fluid relaxation time, $\eta_s$ and $\eta_p$ are the solvent and polymer viscosities respectively, both of which are constant in these models (for the UCM model the solvent contribution is removed, $\eta_s=0$).

In order to increase the stability of the numerical method, we used the log-conformation procedure [28], in which we solve for $\boldsymbol{\Theta} = \log(\mathbf{A})$,

$$\frac{\partial \Theta}{\partial t} + (\mathbf{u} \cdot \nabla)\Theta - (\mathbf{R}\Theta - \Theta\mathbf{R}) - 2\mathbf{E} = \frac{1}{\lambda}\left(e^{-\Theta} - \mathbf{I}\right) \qquad (4)$$

where $\mathbf{A}$ is the conformation tensor, which can be related to the extra-stress tensor as $\mathbf{A} = \frac{\lambda}{\eta_p}\boldsymbol{\tau}_p + \mathbf{I}$, $\mathbf{E}$ and $\mathbf{R}$ are the traceless extensional component and the pure rotational component of the velocity gradient tensor, $\left(\nabla \mathbf{u}^T\right)_{ij} = \left(\partial u_i / \partial x_j\right)$, and $\mathbf{I}$ is the identity matrix [28,29].

An implicit finite-volume method was used to solve the governing equations. The method is based on a time-marching pressure-correction algorithm formulated with the collocated variable arrangement as originally described in Oliveira et al. [30] with subsequent improvements documented in Alves et al. [31]. The interested reader is referred to Afonso et al. [29] for more details and the corresponding numerical implementation and we only give a succinct overview here to avoid unnecessary repetition. The governing equations are transformed first to a generalized (usually non-orthogonal) coordinate system but the Cartesian velocity and stress components are retained. The equations are subsequently integrated in space over control volumes (cells with volume $V_P$) forming the computational mesh, and in time over a time step ($\delta t$), so that sets of linearized algebraic equations are obtained, having the general form:

$$a_P \phi_P = \sum_F a_F \phi_F + S_\phi, \qquad (5)$$

to be solved for the velocity components and for the logarithm of the conformation tensor. In these equations $a_F$ are coefficients accounting for advection and diffusion in the momentum equation and advection for the logarithm of the conformation tensor equations. The source term $S_\phi$ is made up of all contributions that are not included in the terms with coefficients. The subscript P denotes the cell under consideration and subscript F its corresponding neighbouring cells. The central coefficient of the momentum equation, $a_P$, is given by

$$a_P = \frac{\rho V_P}{\delta t} + \sum_F a_F, \qquad (6)$$

while for the log-conformation tensor equation is given by

$$a_P = \frac{\lambda V_P}{\delta t} + \sum_F a_F^\theta \qquad (7)$$

where $a_F^\theta$ contains only the convective fluxes multiplied by $\lambda/\rho$.

After assembling all coefficients and source terms, the linear sets of equations (5) are solved initially for the logarithm of the conformation tensor and subsequently for the Cartesian velocity components. In general, these newly-computed velocity components do not satisfy continuity and therefore need to be corrected by an adjustment of the pressure differences which drive them. This adjustment is accomplished by means of a pressure-correction field obtained from a Poisson pressure equation according to the SIMPLEC algorithm [32] to obtain a velocity field satisfying continuity. To discretize the convective fluxes, the method uses the CUBISTA high-resolution scheme, especially designed for differential constitutive equations [31]. In the current study our interest is restricted to creeping-flow conditions (i.e. $Re \rightarrow 0$) in which case the advection terms of the momentum equation (i.e. the second term on the left side of Eq. (2)) are neglected. For the time-step discretization an implicit first-order Euler method was used, since we are primarily interested in the steady-state solution. We note that when steady-state conditions are achieved, the transient term used in the momentum equation ($\partial \mathbf{u}/\partial t$) for time marching vanishes, and we recover the Stokes flow equation for creeping flow.

## 3   Geometry, computational meshes and boundary conditions

The lid-driven cavity is shown schematically in **Figure 1**. To investigate the role of aspect ratio ($\Lambda=H/L$) we have modelled eight different geometries ($\Lambda$=0.125, 0.25, 0.5, 1.0, 1.5, 2.0, 3.0 and 4.0): thus "tall" enclosures correspond to aspect ratios greater than one and squat (or "shallow") enclosures to aspect ratios less than one. For each aspect ratio three consistently refined meshes have been used to enable the estimation of the numerical uncertainty of the results: for the square cavity additionally, a fourth finer mesh is used. For each geometry the central core of the cavity ($0.1 \leq x/L \leq 0.9$, $0.1 \leq y/H \leq 0.9$) is covered with a uniform mesh which is progressively refined outside of this core region in both *x* and *y* directions so that the minimum cell size occurs in the four corners of the geometry. By construction, each mesh is symmetric about both the vertical and horizontal centrelines. Each mesh has an odd number of cells in both directions so that the variables are calculated exactly along the centrelines. The refinement procedure consists of halving the size of cells in both directions (and reducing the cell expansion/contraction factors accordingly) thus the total number of cells increases by essentially a factor of four between two meshes (to ensure an odd number of cells in each mesh the increase is not exactly a factor of four). The main characteristics of the meshes used for each aspect ratio are given in **Table 2**, including the total number cells (NC) and the minimum cell spacing which occurs at the corners ($\Delta x_{min}/L$ and $\Delta y_{min}/L$).

The boundary conditions applied to the three stationary walls are no slip and impermeability (i.e. $u = v = 0$, where $u$ and $v$ are the Cartesian components of the velocity vector). For the moving wall, as discussed in the Introduction, the unregularized (R0) lid-velocity distribution *viz*.

$$\text{R0:} \qquad u(x) = U, \qquad (8)$$

gives major numerical issues at the corners for viscoelastic fluids, due to the localized infinite acceleration applied to the fluid. The local extensional rate d$u$/d$x$ is infinite, theoretically, and the classical viscoelastic models here considered develop infinite stresses. Even though the numerical approximation introduces some degree of local smoothing, the method is unable to cope with the local stress peaks developed at the corners and fail to give a converged iterative solution. Hence, using this unregularised profile (which we shall refer to as "R0" henceforth) we could obtain converged solutions only in the case of low Weissenberg numbers (i.e. essentially Newtonian fluids only, for $Wi > 0.02$ there are already noticeable oscillations of the computed strength of the main recirculation). We note that this is in marked contrast with the results of Yapici et al. [21] who obtained results up to $Wi$=1.0 for an unregularised profile. One common way of regularising the lid-velocity is to use a polynomial function

$$\text{R1:} \quad u(x) = 16U(x/L)^2(1 - x/L)^2, \qquad (9)$$

such that both the velocity and the velocity gradient vanish at the corners [19,20]. The use of such a regularization significantly reduces the strength of the main recirculation region within the cavity ($|\psi_{min}|$ decreases in the Newtonian $\Lambda= 1$ case by about 16% for example [33]) and so to better mimic the unregularised idealised problem we also investigated the use of two weaker forms of regularization such that the velocity is uniform over the middle 60% of the moving wall

$$\text{R2:} \quad \begin{cases} u(x) = [1/(0.2^2 0.8^2)]U(x/L)^2(1-x/L)^2 & 0 \le x/L \le 0.2 \\ u(x) = U & 0.2 < x/L < 0.8, \\ u(x) = [1/(0.2^2 0.8^2)]U(x/L)^2(1-x/L)^2 & 0.8 \le x/L \le 1 \end{cases} \quad (10)$$

and over 80% of its length

$$\text{R3:} \quad \begin{cases} u(x) = [1/(0.1^2 0.9^2)]U(x/L)^2(1-x/L)^2 & 0 \le x/L \le 0.1 \\ u(x) = U & 0.1 < x/L < 0.9 \,. \\ u(x) = [1/(0.1^2 0.9^2)]U(x/L)^2(1-x/L)^2 & 0.9 \le x/L \le 1 \end{cases} \quad (11)$$

Note that the velocity and velocity gradient also vanish at the corners for regularizations R2 and R3. However, although the velocity profile is continuous, the velocity gradient is not continuous at the points of change between the polynomial and the constant velocity profile, for R2 and R3. The different wall velocity profiles (i.e. R0, R1, R2 and R3), normalised using the peak velocity $U$, are shown in **Figure 2**. It is this *peak* velocity that is used as a characteristic velocity scale in our Deborah and Weissenberg number definitions. The average dimensionless lid velocity (i.e. $\overline{U} = \int_0^L u \, dx/L$) for each regularization is $0.533U$ (R1), $0.751U$ (R2) and $0.870U$ (R3). Finally it is important to highlight that these regularized velocity profiles introduce a natural modification to the estimate of a characteristic acceleration/deceleration time of the flow. For R1, the velocity increases from zero to $U$ over a distance ($L^*$) of $0.5L$, for R2 this decreases to $0.2L$ and for R3 to $0.1L$. Using the average velocity over distance $L^*$, $U^*_m = \int_0^{L^*} u \, dx/L^*$, a modified Deborah number $De^* = \lambda U^*_m / L^*$ can be determined such that $De^*_{R1} = 1.067\, De_{R1}$, $De^*_{R2} = 1.885\, De_{R2}$ and $De^*_{R3} = 3.523\, De_{R3}$. It is worth noting that although theoretically $De^*_{R0} \to \infty$, meaning that obtaining a steady-state solution should not be possible, numerically $De^*$ depends on the mesh resolution (since the velocity jumps from zero to $U$ over a finite distance $dx$). For example for mesh M4, $\Lambda=1$, $De_{R0} = 0.01$ corresponds to a $De^*_{R0} = 16.7$. Interestingly, the numerical difficulties, first seen as oscillations in the convergence trend of the residuals, appear once $De^* \approx \mathcal{O}(1)$.

## 4  Comparison with literature results and numerical accuracy

The numerical investigation of eight different aspect ratios – using three different lid-velocity regularizations – for viscoelastic fluids over a range of Deborah (or Weissenberg) numbers, even in the limit of creeping flow, results in a large data set (approximately 600 simulations). Therefore, in this section only some representative data, which highlight typical levels of uncertainty, are presented.

Comparison of our data for Newtonian fluids with results in the literature (the square cavity case, R0), presented in **Table 3,** shows excellent agreement. The minimum stream function value (i.e. the volumetric rate per unit depth of flow induced in the main recirculation region) agrees with values in the literature to within 0.05%. The minimum $u$ velocity along $x/L$=0.5 also agrees to literature results within 0.05%. There is a mild discrepancy (~1%) with the results of Sahin and Owens [4] in the minimum value of $v$ computed along $y/H$=0.5, but this may be a consequence of their "leaky" boundary conditions near the corners. The agreement of this quantity with the results of Yapici et al. [24] is, as for the other quantities, better than 0.1%.

Comparison of our data for viscoelastic fluids, in this case for the Oldroyd-B model with a solvent-to-total viscosity ratio ($\beta = \eta_s / (\eta_s + \eta_p)$) of 0.5 and $De$=0.5 and 1.0 (square cavity case R1), shown in **Table 4**, indicates that, at $De$=0.5, our results are in good agreement with Pan et al. [19]. At higher levels of elasticity, $De$=1.0, the large normal stresses generated reveal a greater degree of sensitivity and non-negligible differences between results in both our finest two meshes and, also, in comparison with the data of Pan et al. [19]. The effects of mesh density on the accuracy of the numerical results are shown in **Tables 5-7** for $\Lambda$=1, 0.125 and 4, respectively. Overall, the difference between the results on the finest mesh and the extrapolated values – obtained using Richardson's technique [34] – are less than 0.2% for these quantities.

## 5. Creeping Newtonian flow

The streamline patterns for Newtonian flow – wall regularization R3 and mesh M3 – are shown in **Figure 3** for low aspect ratio ($\Lambda < 1$, **Figure 3(a)**) and high aspect ratio ($\Lambda \geq 1$ **Figure 3(b)**). As discussed in the Introduction (and in [2]), for high aspect ratios the streamlines essentially collapse in the top region of the cavities and the maximum absolute value of the stream function – the variation of which with aspect ratio is shown in **Figure 4** – becomes independent of aspect ratio for $\Lambda \geq 1$. For small aspect ratios the scenario is more complex although the asymptotic limit when $\Lambda \to 0$ does allow analytical expressions for the maximum stream function, velocity and stress components to be derived (presented in Appendix A). Under these assumptions the maximum absolute value of stream function is expected to vary linearly with the aspect ratio as $\dfrac{|\psi_{min}|}{UL} = \dfrac{4}{27}\Lambda$, and this linear relationship is also included in **Figure 4**. Excellent agreement between this analytical solution and computations can be seen with aspect ratios up to $\Lambda=0.5$, especially for the unregularized and weakly regularized wall velocities (R0 and R3). In addition, the intersection between that linear variation and the value $|\psi_{min}/UL|=0.1$ at high aspect ratio shows that the "tall" cavities start at $\Lambda \approx 0.7$.

The effects of wall regularization are subtle, yet important. In the square cavity case, $\Lambda=1$, regularizing using the standard polynomial function [18, 19] (R1 in our nomenclature) reduces quantitatively the strength of the main recirculation, by about 16%, (in agreement with previous studies [33]) but the qualitative effect on the streamlines (**Figure 5**) appears to be minor except close to corners A and B, as might be expected. At the lower aspect ratios, however, there are significant qualitative differences between the streamline patterns and the unregularized streamlines are essentially straight over the middle 75% of the cavity. Changing the regularization such that it better approximates the unregularized case, e.g. R3, essentially increases the vortex strength back to its unregularized value (see **Table 5** for example) and better captures the streamline patterns (although close to the corners A and B differences are still apparent – **Figure 5**). To better illustrate these effects, in **Figure 6** we plot contours of the flow type classifier. The flow-type parameter $\xi$ is used to classify the flow locally, and here we use the criterion proposed by Lee et al. [35], $\xi \equiv \dfrac{|\mathbf{D}|-|\boldsymbol{\Omega}|}{|\mathbf{D}|+|\boldsymbol{\Omega}|}$; where $|\mathbf{D}|$ and $|\boldsymbol{\Omega}|$ represent the magnitudes of

the rate of deformation tensor and vorticity tensor, $|\mathbf{D}| = \sqrt{\frac{1}{2}(\mathbf{D}:\mathbf{D})}$ and $|\mathbf{\Omega}| = \sqrt{\frac{1}{2}(\mathbf{\Omega}:\mathbf{\Omega})}$. As such, $\xi = 1$ corresponds to pure extensional flow, $\xi = 0$ corresponds to pure shear flow and $\xi = -1$ corresponds to solid-body rotation flow.

As seen in **Figure 6**, the regularization of the lid velocity significantly affects the flow close to the lid ends. Without regularization (R0), the flow close to the wall is mainly shear dominated. However, the wall regularization induces a strong component of extensional flow close to corners A and B due to the acceleration and deceleration of the fluid at these corners. As this acceleration region decreases (R1→ R2 → R3) the region of extensional-dominated flow also decreases and approaches the "true" lid-driven cavity flow field (R0).

Given the basic modification to the flow field induced by the regularization classically used for viscoelastic fluids [18, 19], care must be taken when comparing regularized simulation results with experimental results [9, 10, 17]. This issue is probably why Grillet et al. [16] implemented a "leaky" boundary condition at corners A and B. In contrast here we attempt to tackle this problem via modification of the classical regularization (R1).

## 6. Viscoelastic flow

### 6.1    Steady-state flow field

The addition of fluid elasticity induces changes to the flow pattern in the lid-driven cavity, and, in particular, a breaking of fore-aft symmetry relative to the $x/L=0.5$ line. **Figure 7** presents the computed streamlines for Newtonian and viscoelastic fluid flow for different aspect ratios, $\Lambda=0.125, 0.25, 0.5$ and 1. For the viscoelastic fluid flow, the cases illustrated correspond to the highest *De* where steady flow is observed with regularization R3 (*De*=0.15). For the Newtonian fluid the flow field is symmetric, as it must be in creeping flow, due to the linearity of the Stokes flow.

The large normal stresses that are generated for the viscoelastic fluid as *De* increases are advected in the downstream direction leading to an increase of the flow resistance, and to compensate for this effect the eye of the recirculation region progressively shifts in the upwind direction - towards corner A as illustrated in **Figure 1** - breaking the symmetry observed for Newtonian fluids. This effect is in excellent agreement with experimental observations for Boger fluids [9]. The increase of the normal stresses with increasing *De*, and concomitant higher flow resistance, induces a decrease in the strength of the main recirculating flow (the flow closest to the lid), i.e. a reduction of $|\psi_{\min}|$ as illustrated in **Figure 8** for different aspect ratios as a function of *De* (for three different lid velocity regularizations). This effect is akin the vortex suppression by elasticity seen in many other flow situations, for example in sudden expansion geometries, and entails the coupling of elastic hoop stresses and curved streamlines (as discussed below in **Section 6.2**). For the lower aspect ratio cases and the regularizations (R2 and R3) closer to the unregularized situation (R0), the streamlines in a large region about the central section of the cavity are straight, with curvature confined to a small region near the lateral walls (see eg. **Figure 4** top), so the mechanism of the vortex suppression should be less effective. Indeed, we find in **Figure 8** a slight initial increase, although small (about 1-2 %) of the vortex intensity with elasticity for the lower curve cases of $\Lambda=0.125$ and $0.25$ with regularizations R2 and R3, which we interpret as resulting from straighter streamlines.

Similarly to the Newtonian fluid flow case, the intensity of recirculating fluid increases with aspect ratio up to $\Lambda=1$, before saturating, and further increases of $\Lambda$ have a negligible effect on $|\psi_{\min}|$ as shown by the data collapse for $\Lambda \geq 1$. The additional recirculation zones that are formed as $\Lambda$ increases, e.g. shown for a Newtonian fluid in **Figure 5**, are significantly weaker, having a negligible effect on $|\psi_{\min}|$. With the regularization R1 (black symbols), since the lid-velocity profile is smoother and has a lower average velocity relative to the R3 regularization, the maximum value of the stream function magnitude is also lower and higher Deborah numbers can be achieved prior to the onset of a purely-elastic instability, which we discuss next.

**6.2 Onset and scaling of a purely-elastic instability**

The critical conditions for the onset of a purely-elastic instability are presented in **Figure 9**, both in terms of a critical reciprocal Weissenberg number ($1/Wi_{\mathrm{cr}}$) and a Deborah number ($De_{\mathrm{cr}}$) as a function of aspect ratio, using three different lid regularizations (R1-R3), computed using mesh

M3. The critical condition is identified when a steady-state solution can no longer be obtained: thus the purely-elastic instability, in all cases examined here, gives rise to a time varying flow in agreement with experimental observations [10]. We confirmed that this critical condition is independent of the time-step used in the time-matching algorithm.

For a given level of wall regularization, the flow instabilities for different aspect ratios occur approximately at a constant Deborah number, e.g. for R1 the values are within $De = 0.625 \pm 0.050$. Consequently, $1/Wi_{cr}$ varies linearly with aspect ratio, with $1/Wi_{cr} = 1.60\Lambda$ for regularization R1. As the regularization is weakened, and the forcing better approximates the "true" unregularized case i.e. R1 → R2 → R3, the critical $De$ (or $Wi$) decreases significantly from 0.63 (R1) to 0.33 (R2) to 0.18 (R3). Thus the precise regularization used can decrease the critical $De$ to about one third. The use of an average wall velocity ($\bar{U} = \int_0^L u \, dx / L$), instead of the peak velocity $U$, reduces these differences in critical value slightly (to about half). The use of a more appropriate characteristic time, based on the distance over which the lid velocity grows, to define $De^*$ and $Wi^*$ as described in **Section 3**, is much better able to collapse the critical conditions as shown in **Figure 9(b)**. For the two weakest forms of regularization (R2 and R3) this practically collapses the critical values. If this scaling is representative of the controlling dynamics it would imply that the unregularized lid is unstable for vanishingly small elasticities (given the infinite acceleration, or zero acceleration time, for a fluid element to go from rest to velocity $U$, or equivalently $De^* \to \infty$), but as shown in **Section 3**, numerically $De^*$ is finite since the mesh elements do not have zero length.

**Figure 10(a)** presents the contours of the Pakdel-McKinley criterion $M_{crit} = \sqrt{(\lambda u / \Re)(\tau_{11}/\eta_0 \dot{\gamma})}$ [8,12], (where $\tau_{11}$ is the tensile stress in the local streamwise direction, $\dot{\gamma}$ is the local shear rate and $\Re$ is the local streamline radius of curvature), for $\Lambda=0.25, 0.5, 1, 2$ and $4$ for the highest steady $De$. The maximum values of $M_{crit}$ are of the same order as observed in previous studies, between 3 and 4 [12], namely 3.6 ($\Lambda=0.25$), 3.9 ($\Lambda=0.5$), 3.5 ($\Lambda=1, 2$ and $4$). These critical $M$ values are located near the downstream corner, where large normal stresses generated on approaching corner B are advected into a region of high streamline curvature.

Finally, contours of the flow type parameter are presented in **Figure 10 (b)**, at the highest stable *De* for which the flow remains steady for Λ=0.125 and 1, for both regularizations R1 and R3. For Λ=1, the flow is mainly rotational close to the main vortex centre, shear dominated close to the lid and highly extensional near the top corners in a direction at 45º and in the lower part of the cavity, albeit here the deformation rates and hence stresses are always modest. For Λ=0.125, the flow is mainly shear dominated close to the lid and bottom wall and highly extensional close to the corners and along a thin strand at $y/H \approx 1/3$.

## 5. Conclusions

Viscoelastic creeping flow in a lid-driven cavity was analysed numerically using a finite-volume numerical methodology in combination with the log-conformation technique. The effects of aspect ratio, strength of elasticity via the Deborah/Weissenberg numbers and, in contrast to previous studies, the effect of regularization type for the lid velocity were investigated. As discussed in the introduction and, as observed in previous studies for Newtonian fluids, the streamlines essentially collapse and the stream function becomes independent of aspect ratio, for Λ≥1. For low aspect ratios, the maximum value of the stream function magnitude agrees with a simple parallel-flow approximation analytical solution.

The effect of elasticity on the steady flow characteristics was elucidated, including the experimentally-observed shift of the main vortex centre in the direction of the upstream corner (A) and the breaking of fore-aft symmetry with increasing elasticity. Increasing the elasticity was also found to reduce the flow strength induced by the lid. The critical conditions for the onset of purely-elastic flow instabilities were characterized by plotting $1/Wi_{cr}$ (and $De_{cr}$) as a function of aspect ratio. The value of the Pakdel- McKinley criterion [12] at the critical conditions was similar to previous studies, between 3 and 4. In accordance with the experimental results of Pakdel and McKinley [8,10,12], we find a linear scaling of reciprocal Weissenberg number with aspect ratio, corresponding to a constant Deborah number, the numerical value of which is dependent on the wall regularization used. By introducing a modified Deborah number, based on an average acceleration time for the flow adjacent to the lid to accelerate from *u*=0 to *u*=*U*, a reasonable collapse for the various Λ is achieved. The flow-type parameter was also computed to characterize the different regions of the lid-driven cavity flow. Further studies are required to investigate more closely the generated time-dependent flows.

**Acknowledgements**. RGS and MAA acknowledge funding from the European Research Council (ERC) under the European Union's Seventh Framework Programme (ERC Grant Agreement n. 307499). RJP acknowledges funding for a "Fellowship in Complex Fluids and Rheology" from the Engineering and Physical Sciences Research Council (EPSRC, UK) under grant number (EP/M025187/1). AMA acknowledges Fundação para a Ciência e a Tecnologia (FCT) for financial support through the scholarship SFRH/BPD/75436/2010. AM acknowledges support from the UK Engineering and Physical Sciences Research Council (EP/I004262/1). Research outputs generated through the EPSRC grant EP/I004262/1 can be found at http://dx.doi.org/10.7488/ds/1330.

**Appendix A: Parallel flow analysis in the small aspect ratio limit**

For small aspect ratio cavities (i.e. $L>>H$) the flow far away from the end walls and close to the vertical centreplane ($x/L \approx 0.5$) should be well approximated by a fully developed Couette-Poiseuille flow (i.e. $v=0$, $dp/dy=0$, $dp/dx=$constant). For the creeping-flow situation considered here the Navier-Stokes equations thus simplify to

$$0 = -\frac{dp}{dx} + \eta_0 \left( \frac{\partial^2 u}{\partial x^2} + \frac{\partial^2 u}{\partial y^2} \right). \tag{A1}$$

Equation (A1) is valid for both Newtonian and UCM/Oldroyd-B fluids due to the constant viscosity of both models and therefore the assumption of a pure-shear flow. In addition, a simple scaling argument can be used to estimate the relative order of magnitude of the two viscous terms

$$\frac{\partial^2 u}{\partial x^2} \approx \frac{U}{L^2} \qquad \frac{\partial^2 u}{\partial y^2} \approx \frac{U}{H^2}, \tag{A2}$$

in the small aspect ratio limit $L>>H$ and therefore

$$\frac{\partial^2 u}{\partial y^2} >> \frac{\partial^2 u}{\partial x^2}. \tag{A3}$$

Solving Eq. (A1) subject to the boundary conditions $y=0$, $u=0$ and $y=H$, $u=U$ gives the velocity distribution $u(y)$ in terms of a constant (proportional to the constant pressure gradient in the streamwise direction)

$$u(y) = C(y^2 - yH) + \frac{Uy}{H}. \tag{A4}$$

Realising that the total volumetric flow rate must be equal to zero under these conditions (i.e. $\int_0^H u(y)dy = 0$) enables us to determine the velocity profile as

$$u(y) = \frac{3U}{H^2}(y^2 - yH) + \frac{Uy}{H}. \tag{A5}$$

The minimum value of the stream function (defined as $u = d\psi/dy$) will occur at a $y$ location where $u=0$ and this occurs at $y=2H/3$. Thus the minimum stream function will be

$$\psi_{min} = -\frac{4}{27}UH. \tag{A6}$$

The stream function can be normalized using $UL$ and introducing the aspect ratio ($\Lambda=H/L$) gives

$$\frac{\psi_{min}}{UL} = -\frac{4}{27}\Lambda. \tag{A7}$$

We can also use the velocity distribution to estimate the maximum dimensionless shear stress on the lid ($=\eta_0(du/dy)_{(y=H)}$)

$$\frac{\tau_{xy}L}{\eta_0 U} = \frac{4}{\Lambda}, \tag{A8}$$

and also the maximum dimensionless axial normal stress

$$\frac{\tau_{xx}L}{\eta_0 U} = \frac{32Wi}{\Lambda}\beta. \tag{A9}$$

# Tables

**Table 1:** Previous numerical studies concerned with lid-driven cavity flow of constant viscosity viscoelastic fluids.

| Reference | Aspect ratios | Constitutive equation | *Wi* | Regularization | Notes |
|---|---|---|---|---|---|
| Grillet et al. [16] | 0.5, 1.0, 3.0 | FENE-CR $L^2$=25, 100, 400 | $\leq 0.24$ | Leakage at corners A and B | FE |
| Fattal and Kupferman [18] | 1.0 | Oldroyd-B, $\beta$=0.5 | 1.0, 2.0, 3.0 and 5.0 | $u(x)=16Ux^2(1-x)^2$ | FD, Log conformation technique |
| Pan et al. [19] | 1.0 | Oldroyd-B, $\beta$=0.5 | 0.5, 1.0 | $u(x)=16Ux^2(1-x)^2$ | FE, Log conformation technique |
| Yapici et al. [24] | 1.0 | Oldroyd-B, $\beta$=0.3 | $\leq 1.0$ | No | FV, First-order upwind |
| Habla et al. [20] | 1.0 | Oldroyd-B, $\beta$=0.5 | $0 \leq 2$ | $u(x,z)=128[1+\tanh 8(t-1/2]x^2(1-x)^2 z^2(1-z^2)$ | FV, 3D, Log conformation technique, CUBISTA |
| Comminal et al. [21] | 1.0 | Oldroyd-B, $\beta$=0.5 | $0.25 \leq 10$ | $u(x)=16Ux^2(1-x)^2$ | FD/FV, Log-conformation, stream function |
| Martins et al. [22] | 1.0 | Oldroyd-B, $\beta$=0.5 | 0.5, 1.0, 2.0 | $u(x)=16Ux^2(1-x)^2$ | FD, Kernel-conformation technique |
| Dalal et al. [23] | 1.0 | Oldroyd-B, $\beta$=0.5 | 1.0 | $u(x)=16Ux^2(1-x)^2$ | FD, Symmetric square root |

FE: Finite-element method; FD: Finite-difference method; FV: Finite-volume method

**Table 2**: Main characteristics of the computational meshes (NC=number of cells)

| Aspect ratio $\Lambda=H/L$ | M1 | | M2 | | M3 | | M4 | |
|---|---|---|---|---|---|---|---|---|
| | $\Delta x_{min}/L = \Delta y_{min}/L$ | NC | $\Delta x_{min}/L = \Delta y_{min}/L$ | NC | $\Delta x_{min}/L = \Delta y_{min}/L$ | NC | $\Delta x_{min}/L = \Delta y_{min}/L$ | NC |
| 0.125 | 0.0012 | 4125 | 0.0006 | 16779 | 0.0003 | 67671 | | |
| 0.25 | 0.0012 | 6765 | 0.0006 | 27307 | 0.0003 | 108405 | | |
| 0.5 | 0.0025 | 3403 | 0.0012 | 13695 | 0.0006 | 54285 | | |
| 1.0 | 0.005 | 1681 | 0.0024 | 6889 | 0.0012 | 27225 | 0.0006 | 108241 |
| 1.5 | 0.005 | 2255 | 0.0024 | 9213 | 0.0012 | 36795 | | |
| 2.0 | 0.005 | 2747 | 0.0024 | 11039 | 0.0012 | 43725 | | |
| 3.0 | 0.005 | 3731 | 0.0024 | 15189 | 0.0012 | 60225 | | |
| 4.0 | 0.005 | 4797 | 0.0024 | 19339 | 0.0012 | 76725 | | |

**Table 3**: Comparison of current results with literature values for Newtonian fluids, for creeping flow, $Re \rightarrow 0$. Unregularized lid (R0), $\Lambda = 1$: minimum values of $u$ computed along $x/L=0.5$; maximum values of $v$ computed along $y/H=0.5$; minimum values of the stream function and the corresponding coordinates $x_{min}/L$, $y_{min}/H$ of the centre of the recirculation.

| Reference | $u_{min}/U$ | $v_{max}/U$ | $\Psi_{min}/UL$ | $x_{min}/L$, $y_{min}/H$ |
|---|---|---|---|---|
| Botella and Peyret [6] | | | -0.100076 | |
| Grillet et al. [16] | | | -0.099931 | 0.5000, 0.7643 |
| Mitsoulis and Zisis [25] | | | -0.0995 | 0.5000, 0.7625 |
| Sahin and Owens [4] | -0.207754 | 0.186273 | -0.100054 | 0.5000, 0.7626 |
| Yapici et al. [24] | -0.207738 | 0.184427 | -0.100072 | 0.5000, 0.7651 |
| Zhang [26] | | | -0.0996 | 0.5000, 0.7645 |
| Habla et al. [20] (*)3D | | | | 0.500, 0.763 (*) |
| Current study M4 | -0.207719 | 0.184425 | -0.100063 | 0.5000, 0.7647 |
| Extrapolated | **-0.207762** | **0.184449** | **-0.100074** | **0.5000, 0.7644** |

**Table 4:** Comparison of current results with literature values for viscoelastic fluids, Oldroyd-B model, $\beta = 0.5$, regularization R1, $\Lambda=1$, $Re=0$.

| Reference | $De$ | MAX($\ln(\tau_{xx}/(\eta_0 U/L))$ at $x=0.5$) | $\Psi_{min}/UL$ | $x_{min}/L$, $y_{min}/H$ |
|---|---|---|---|---|
| Pan et al. [19] | 0.5 | ≈5.5 | -0.0700056 | 0.469, 0.798 |
| Current work M3 | 0.5 | 5.34 | -0.0697524 | 0.464, 0.797 |
| Current work M4 | 0.5 | 5.51 | -0.0697717 | 0.466, 0.800 |
| Current work extr. | 0.5 | 5.57 | -0.0697781 | 0.467, 0.801 |
| | | | | |
| Pan et al. [19] | 1.0 | ≈8.6 | -0.0638341 | 0.439, 0.816 |
| Current work M3 | 1.0 | 7.22 | -0.0618784 | 0.433, 0.821 |
| Current work M4 | 1.0 | 7.80 | -0.0619160 | 0.434, 0.816 |
| Current work extr. | 1.0 | 7.99 | -0.0619285 | 0.434, 0.814 |
| Dalal et al. [23] ( $Re = 1.0$) | 1.0 | - | -0.06141 | 0.432, 0.818 |

**Table 5**: Effect of mesh refinement on: minimum values of $u$ computed along $x/L=0.5$; maximum values of $v$ computed along $y/H=0.5$; minimum values of the stream function for $\Lambda =1$.

|  | Reg | $u_{min}/U$ | $v_{max}/U$ | $\Psi_{min}/UL$ |
|---|---|---|---|---|
| Newtonian M1 | R0 | -0.205678 | 0.183351 | -9.93048x10$^{-2}$ |
| M2 | R0 | -0.207204 | 0.184001 | -9.98509x10$^{-2}$ |
| M3 | R0 | -0.207589 | 0.184353 | -1.00028x10$^{-1}$ |
| M4 | R0 | -0.207719 | 0.184425 | -1.00063x10$^{-1}$ |
| Extrapolated |  | **-0.207762** | **0.184449** | **-1.00074x10$^{-1}$** |
|  |  |  |  |  |
| Newtonian M1 | R3 | -0.205654 | 0.183584 | -9.93471x10$^{-2}$ |
| M2 | R3 | -0.207129 | 0.184158 | -9.98783x10$^{-2}$ |
| M3 | R3 | -0.207494 | 0.184496 | -1.00051x10$^{-1}$ |
| M4 | R3 | -0.207621 | 0.184566 | -1.00084x10$^{-1}$ |
| Extrapolated |  | **-0.207663** | **0.184589** | **-1.00095x10$^{-1}$** |
|  |  |  |  |  |
| UCM $De$=0.1 M1 | R3 | -0.198595 | 0.176940 | -9.67456x10$^{-2}$ |
| M2 | R3 | -0.200029 | 0.177549 | -9.72368x10$^{-2}$ |
| M3 | R3 | -0.200508 | 0.177916 | -9.73744x10$^{-2}$ |
| M4 | R3 | -0.200609 | 0.177995 | -9.74221x10$^{-2}$ |
| Extrapolated |  | **-0.200643** | **0.178021** | **-9.74380x10$^{-2}$** |
|  |  |  |  |  |
| Newtonian M1 | R1 | -0.167654 | 0.146202 | -8.31277x10$^{-2}$ |
| M2 | R1 | -0.168682 | 0.146625 | -8.35121x10$^{-2}$ |
| M3 | R1 | -0.168842 | 0.146698 | -8.36359x10$^{-2}$ |
| M4 | R1 | -0.168885 | 0.146725 | -8.36574x10$^{-2}$ |
| Extrapolated |  | **-0.168899** | **0.146735** | **-8.36646x10$^{-2}$** |
|  |  |  |  |  |
| UCM $De$=0.4 M1 | R1 | -0.116698 | 0.104745 | -5.96625x10$^{-2}$ |
| M2 | R1 | -0.117251 | 0.105013 | -5.99104x10$^{-2}$ |
| M3 | R1 | -0.117398 | 0.105118 | -5.99880x10$^{-2}$ |
| M4 | R1 | -0.117412 | 0.105109 | -5.99925x10$^{-2}$ |
| Extrapolated |  | **-0.117417** | **0.105106** | **-5.99940x10$^{-2}$** |

**Table 6**: Effect of mesh refinement on: minimum values of *u* computed along *x/L*=0.5; maximum values of *v* computed along *y/H*=0.5; minimum values of the stream function for $\Lambda$ =0.125.

|  | Reg | $u_{min}/U$ | $v_{max}/U$ | $\Psi_{min}/UL$ |
|---|---|---|---|---|
| Newtonian M1 | R0 | -0.329378 | 0.235738 | -1.85513x10$^{-2}$ |
| M2 | R0 | -0.332881 | 0.236461 | -1.85523x10$^{-2}$ |
| M3 | R0 | -0.333262 | 0.236633 | -1.85587x10$^{-2}$ |
| Extrapolated | | **-0.333389** | **0.236690** | **-1.85609x10$^{-2}$** |
| | | | | |
| Newtonian M1 | R3 | -0.329378 | 0.156131 | -1.85462x10$^{-2}$ |
| M2 | R3 | -0.332880 | 0.155828 | -1.85455x10$^{-2}$ |
| M3 | R3 | -0.333262 | 0.155765 | -1.85512x10$^{-2}$ |
| Extrapolated | | **-0.333390** | **0.155744** | **-1.85531x10$^{-2}$** |
| | | | | |
| UCM *De*=0.100 | | | | |
| M1 | R3 | -0.329859 | 0.127401 | -1.87031x10$^{-2}$ |
| M2 | R3 | -0.333510 | 0.126345 | -1.87056x10$^{-2}$ |
| M3 | R3 | -0.333959 | 0.126117 | -1.87112x10$^{-2}$ |
| Extrapolated | | **-0.334109** | **0.126040** | **-1.87131x10$^{-2}$** |
| | | | | |
| Newtonian M1 | R1 | -0.325547 | 0.047138 | -1.83536x10$^{-2}$ |
| M2 | R1 | -0.329202 | 0.047014 | -1.83572x10$^{-2}$ |
| M3 | R1 | -0.329548 | 0.046985 | -1.83584x10$^{-2}$ |
| Extrapolated | | **-0.329663** | **0.046976** | **-1.83589x10$^{-2}$** |
| | | | | |
| UCM *De*=0.360 | | | | |
| M1 | R1 | -0.255973 | 0.045105 | -1.48218x10$^{-2}$ |
| M2 | R1 | -0.256215 | 0.044898 | -1.48783x10$^{-2}$ |
| M3 | R1 | -0.256886 | 0.044921 | -1.48948x10$^{-2}$ |
| Extrapolated | | **-0.257110** | **0.0449292** | **-1.49000x10$^{-2}$** |

**Table 7**: Effect of mesh refinement on: minimum values of *u* computed along *x/L*=0.5; maximum values of *v* computed along *y/H*=0.5; minimum values of the stream function for $\Lambda$ =4.

|  | Reg | $u_{min}/U$ | $v_{max}/U$ | $\Psi_{min}/UL$ |
|---|---|---|---|---|
| Newtonian M1 | R0 | -0.192048 | $3.80498 \times 10^{-4}$ | $-9.98683 \times 10^{-2}$ |
| M2 | R0 | -0.194012 | $3.82099 \times 10^{-4}$ | $-1.00644 \times 10^{-1}$ |
| M3 | R0 | -0.194708 | $3.82390 \times 10^{-4}$ | $-1.00817 \times 10^{-1}$ |
| Extrapolated |  | **-0.194940** | **$3.82487 \times 10^{-4}$** | **$-1.00874 \times 10^{-1}$** |
| Newtonian M1 | R3 | -0.191946 | $3.81280 \times 10^{-4}$ | $-9.99168 \times 10^{-2}$ |
| M2 | R3 | -0.193858 | $3.82780 \times 10^{-4}$ | $-1.00673 \times 10^{-1}$ |
| M3 | R3 | -0.194552 | $3.83055 \times 10^{-4}$ | $-1.00843 \times 10^{-1}$ |
| Extrapolated |  | **-0.194782** | **$3.83147 \times 10^{-4}$** | **$-1.00900 \times 10^{-1}$** |
| UCM *De*=0.12 M1 | R3 | -0.183001 | $3.61449 \times 10^{-4}$ | $-9.59420 \times 10^{-2}$ |
| M2 | R3 | -0.184866 | $3.62997 \times 10^{-4}$ | $-9.68424 \times 10^{-2}$ |
| M3 | R3 | -0.185434 | $3.63281 \times 10^{-4}$ | $-9.69867 \times 10^{-2}$ |
| Extrapolated |  | **-0.185623** | **$3.63375 \times 10^{-4}$** | **$-9.70348 \times 10^{-2}$** |

**Figures**

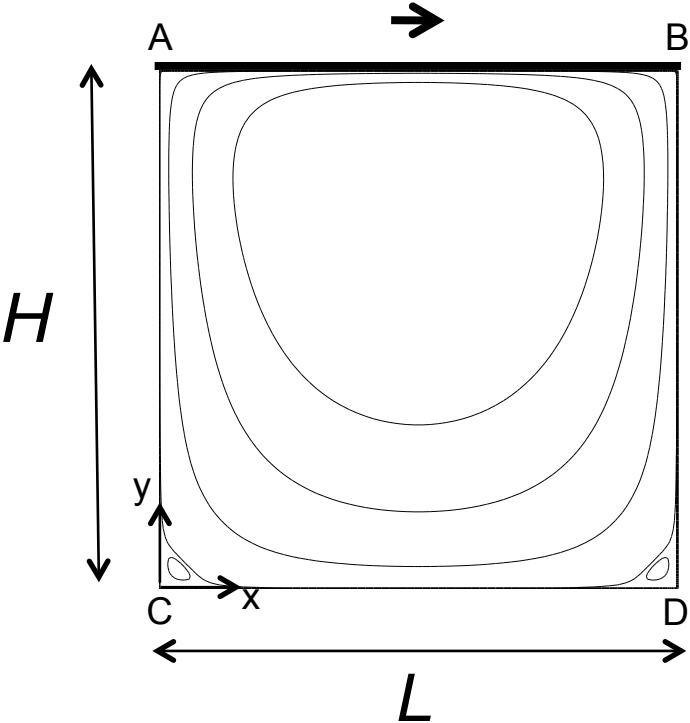

**Figure 1:** Schematic of lid-driven cavity (including representative streamlines for creeping Newtonian flow).

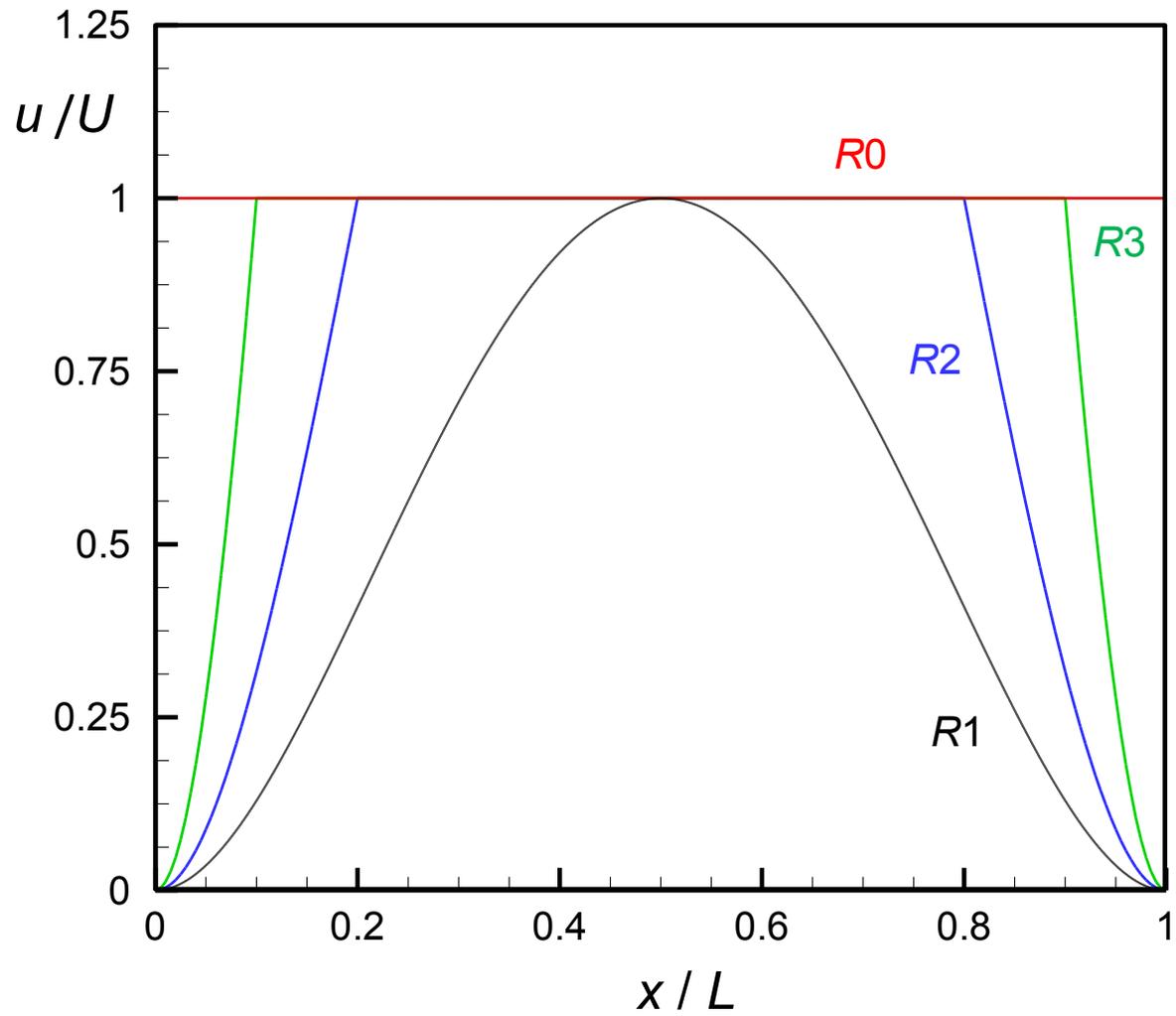

**Figure 2:** Lid velocity profiles for different regularizations used in this work. The average velocities $U^*m$ (from $x=0$ to $x=L^*$, where $u=U$) for each regularization are 0.533 for R1, 0.377 for R2 and 0.352 for R3.

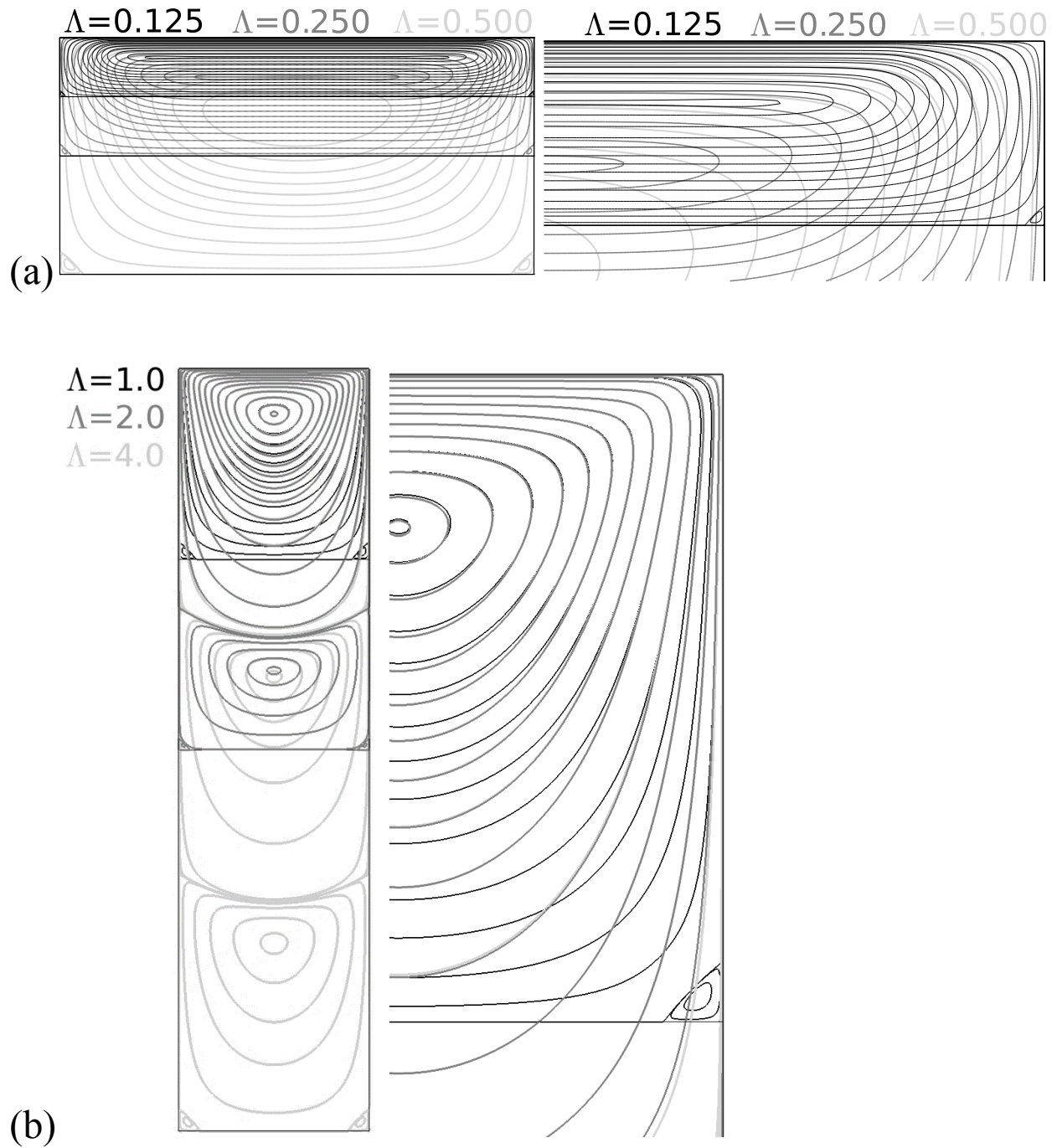

**Figure 3:** Streamlines for Newtonian fluid flow using lid velocity regularization R3 and mesh M3 (a) small aspect ratios ($\Lambda < 1$) (and zoomed region near the downstream corner) (b) large aspect ratios ($\Lambda \geq 1$) (and zoomed region near downstream corner).

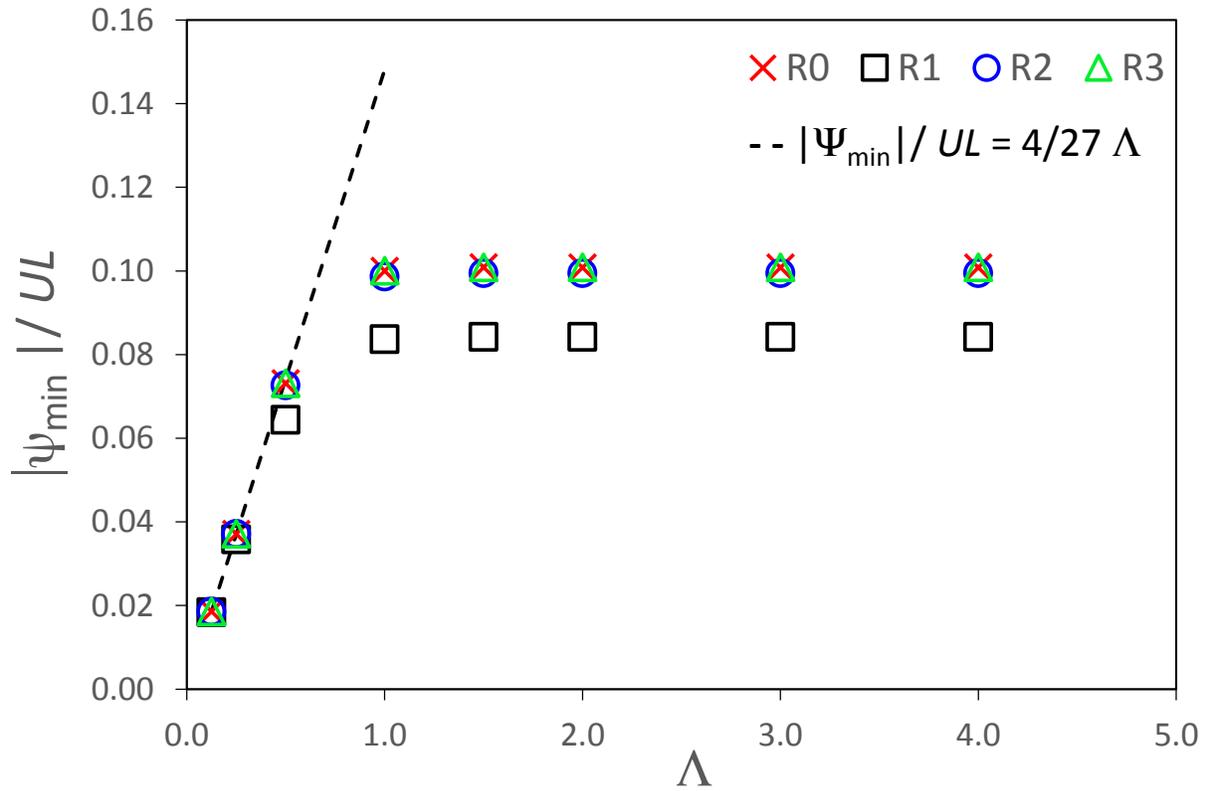

**Figure 4:** Variation of the absolute value of the minimum normalised stream function with aspect ratio for Newtonian fluid flow (regularizations R0, R1, R2 and R3) including the analytical solution for the small aspect ratio limit.

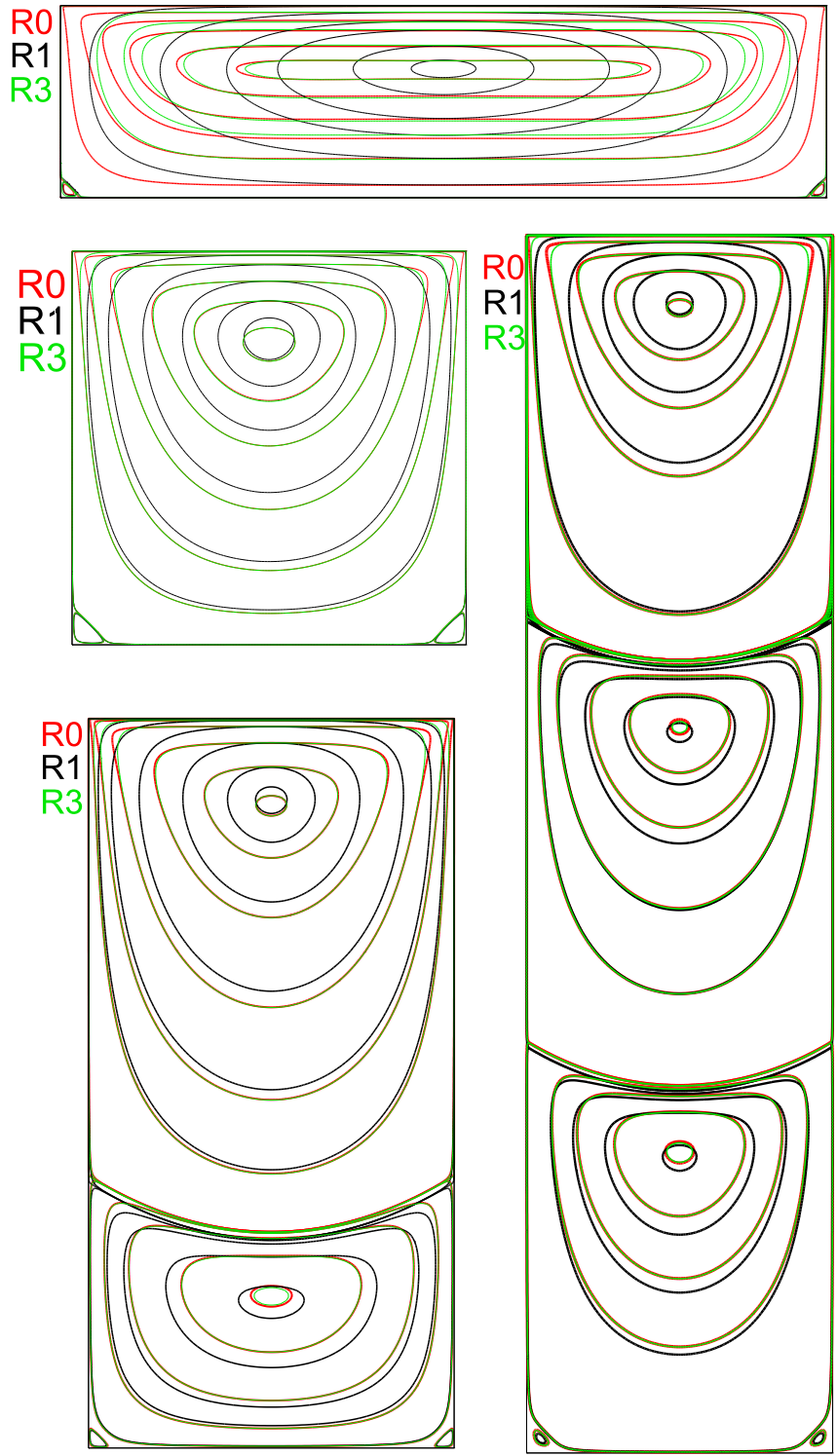

**Figure 5:** Effect of lid velocity regularization (R0, R1, R3) on the computed streamlines for aspect ratios $\Lambda = 0.25$, 1, 2 and 4, using mesh M3.

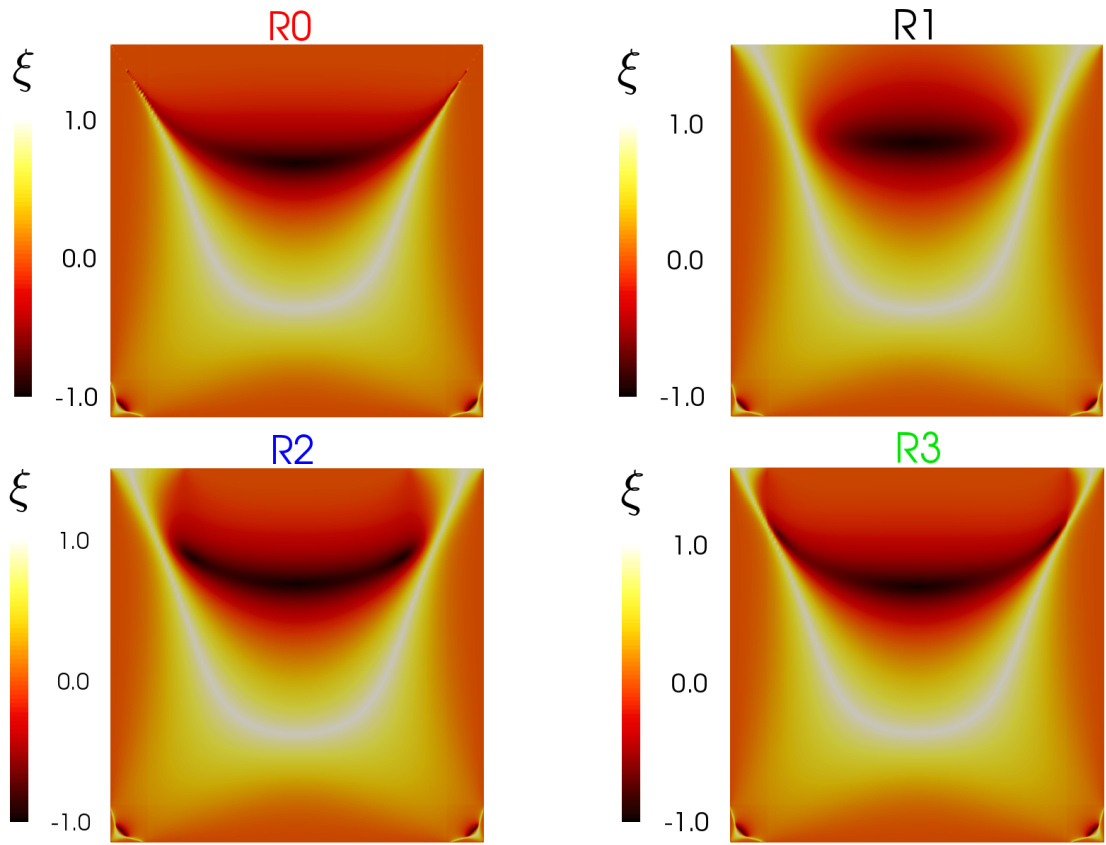

**Figure 6**: Flow type parameter for different lid velocity regularizations for Λ=1 and creeping flow of Newtonian fluids.

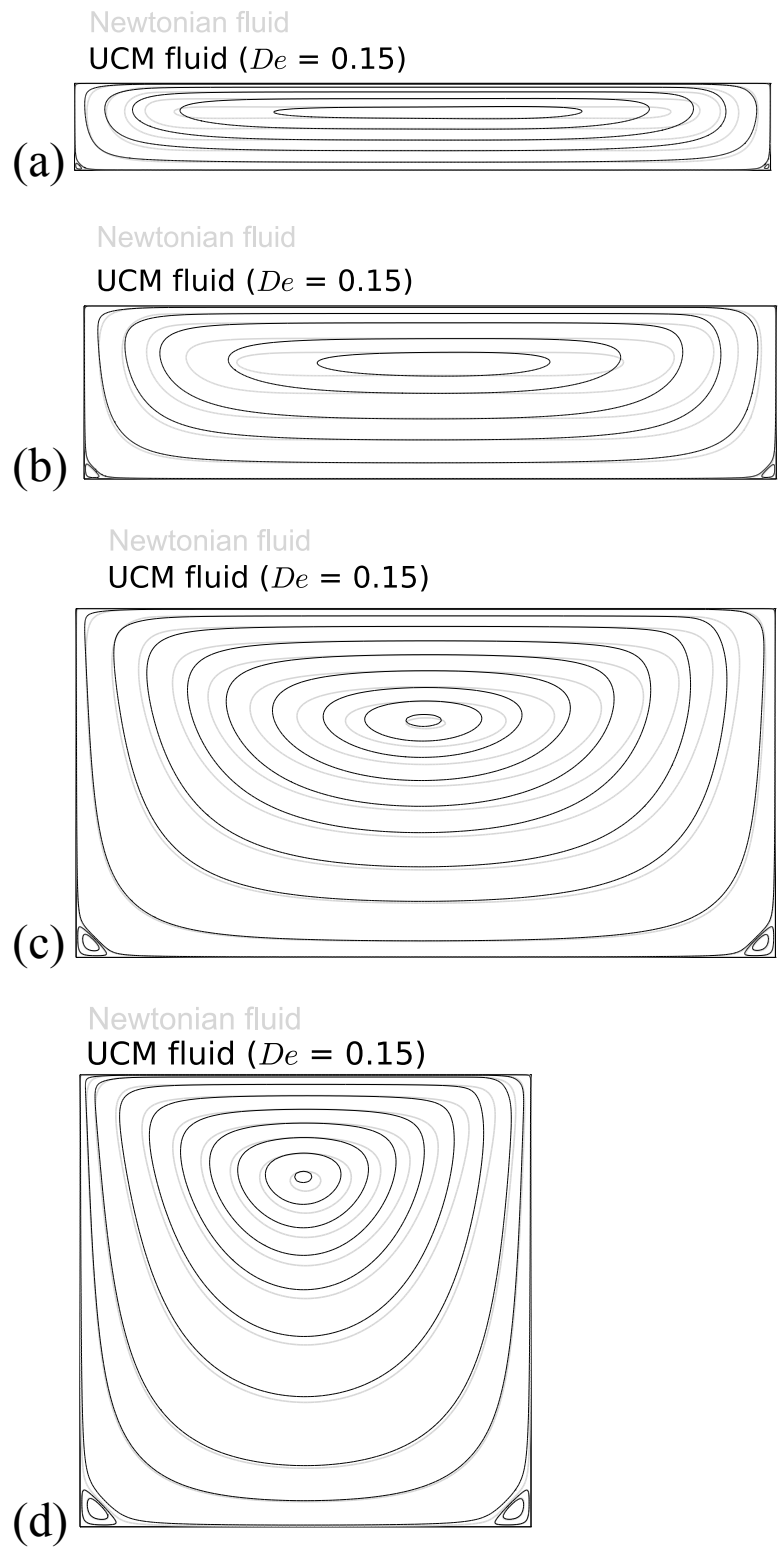

**Figure 7**: Effect of elasticity on streamlines using lid velocity regularization R3 and mesh M3 (a) Λ=0.125 (b) Λ=0.25, (c) Λ=0.50, and (d) Λ=1.00 .

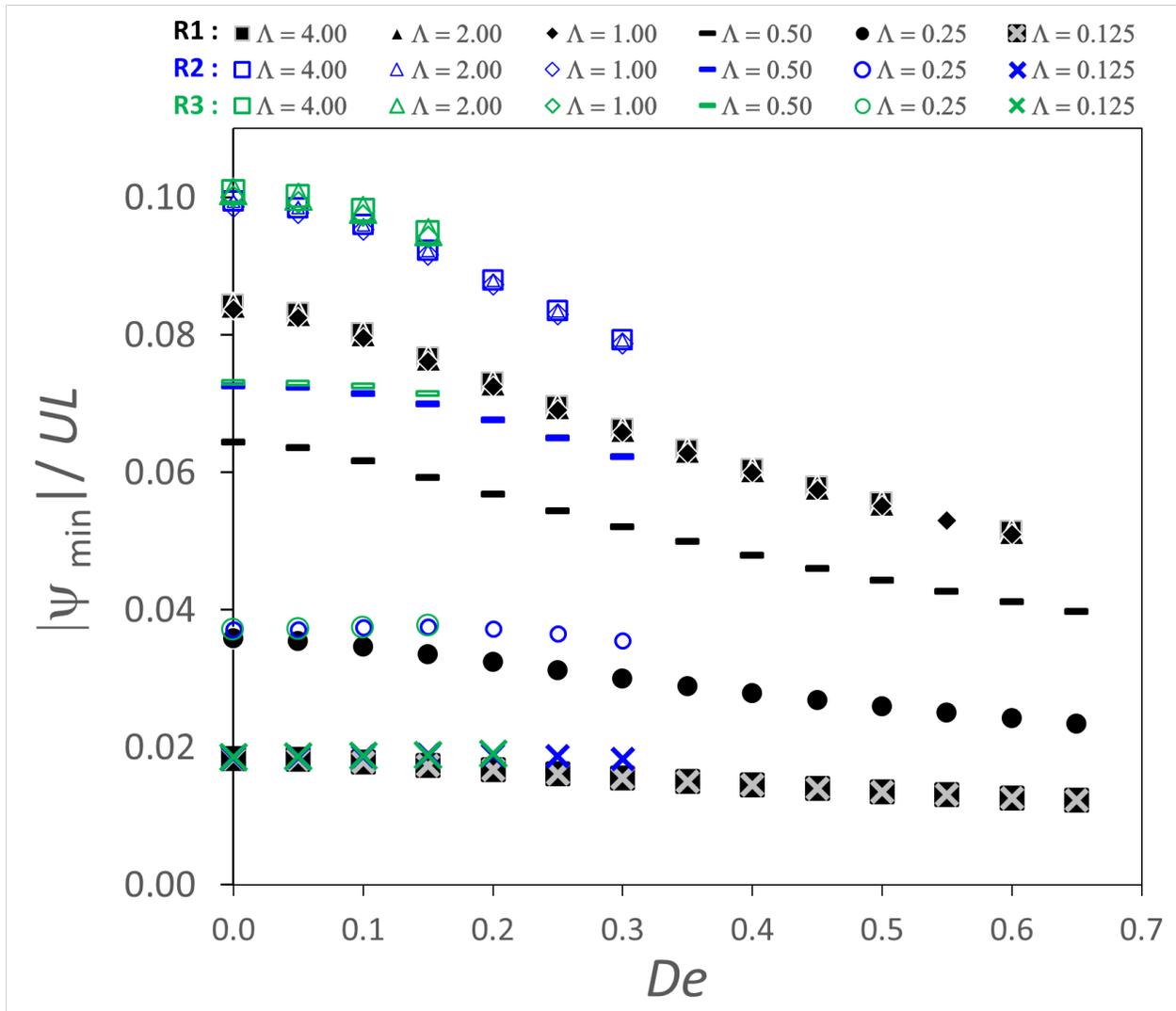

**Figure 8:** Variation of absolute value of minimum dimensionless stream function with *De* for various aspect ratios ($\Lambda$ = 0.125, 0.25, 0.50, 1.00, 2.00, 4.00, with Mesh M3) and lid velocity regularization (R1, R2 and R3).

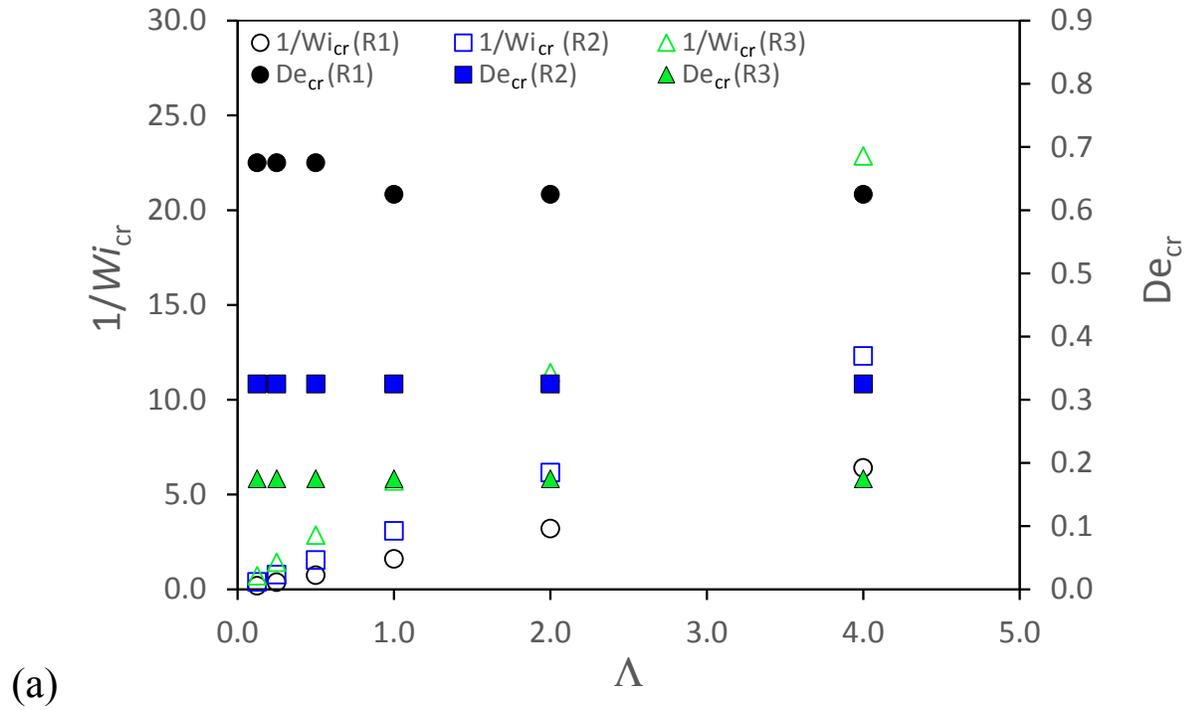

(a)

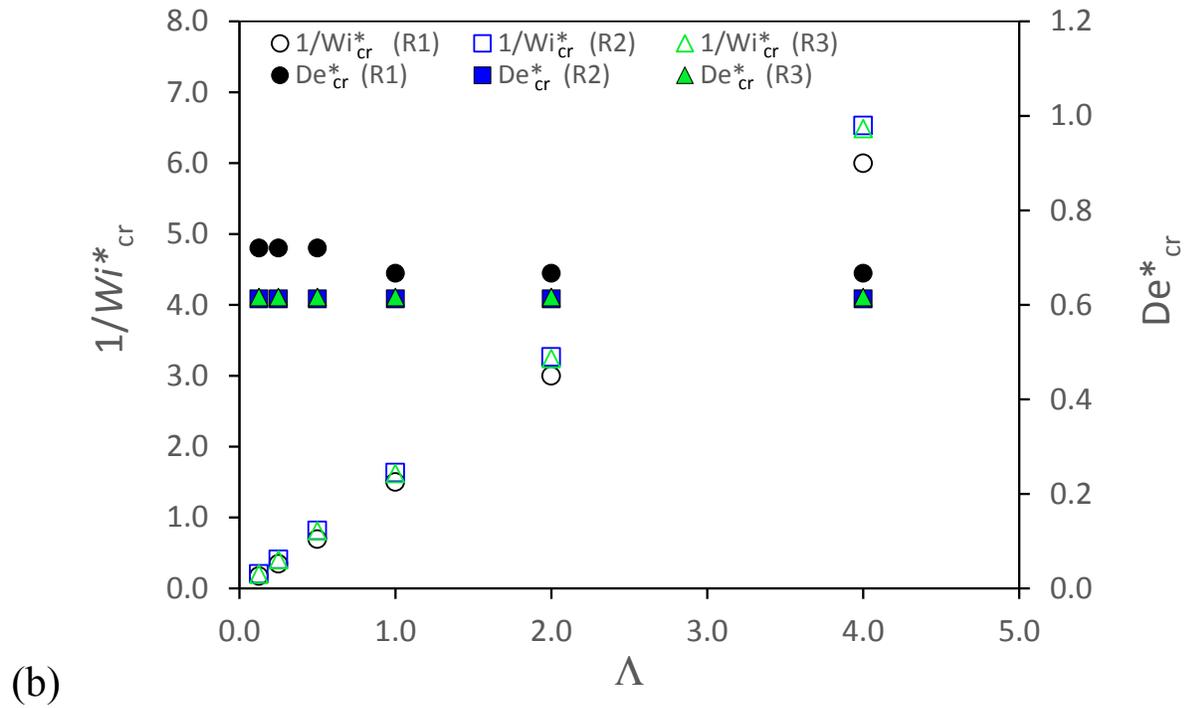

(b)

**Figure 9:** Critical conditions for the onset of purely-elastic instability for different lid velocity regularizations computed in mesh M3: (a) $1/Wi_{cr}$ and $De_{cr}$ versus aspect ratio; (b) $1/Wi^*_{cr}$ and $De^*_{cr}$ versus aspect ratio.

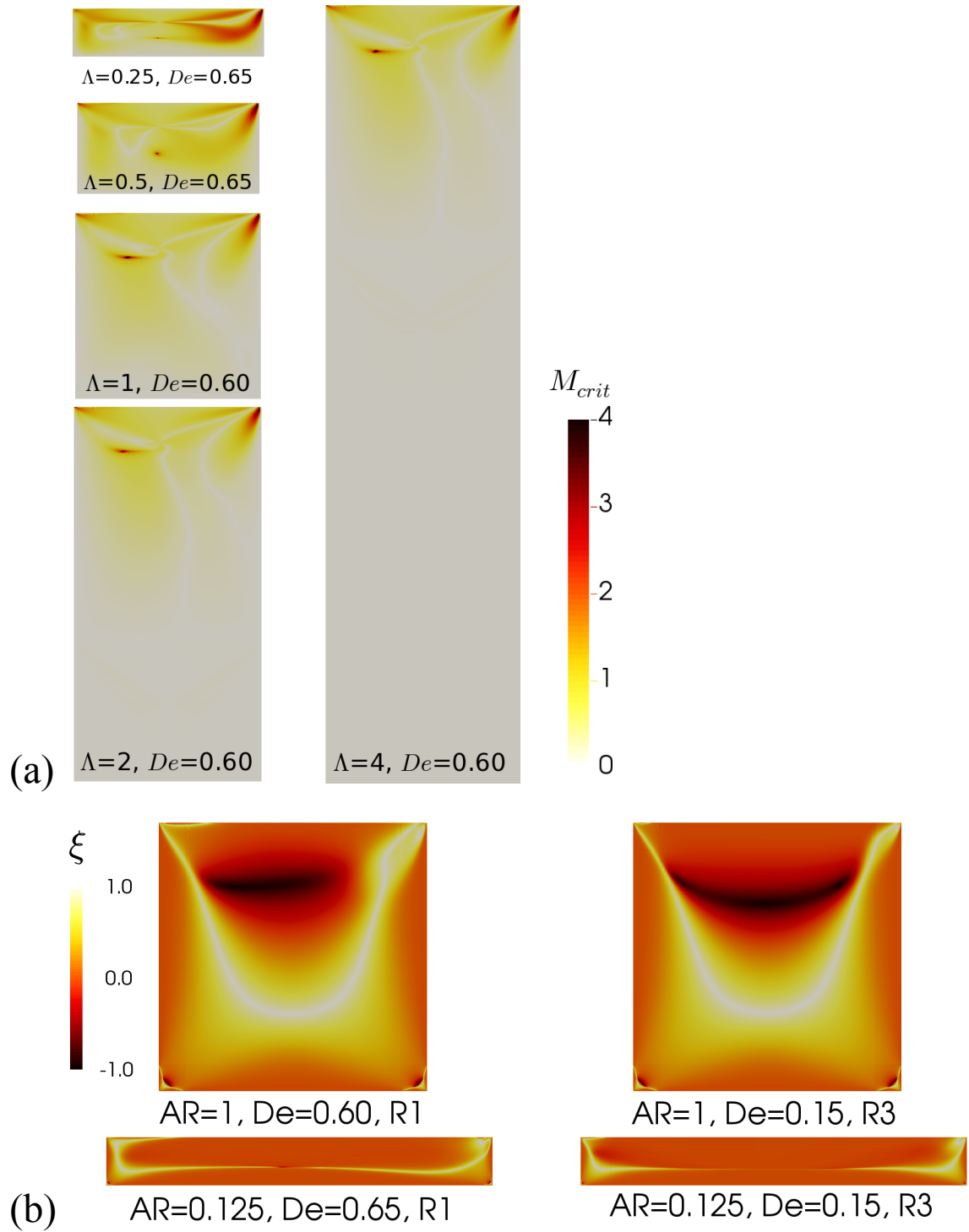

**Figure 10:** Nature of instability shown by contours of: (a) *M* parameter at critical *De*, for regularization R1; (b) Flow type parameter $\xi$ at critical *De* for $\Lambda$=0.125 and 1.0, regularizations R1 and R3.